\documentclass[12pt]{article}
\input epsf.sty
\usepackage{amsmath}
\usepackage{cite}
\usepackage{amssymb}
\usepackage{mathtools}
\usepackage{mathrsfs}
\usepackage{bm}
\usepackage{graphicx}
\usepackage{mathrsfs}
\usepackage{amsmath}
\usepackage{mathrsfs}
\usepackage{amssymb}
\usepackage{color}
\usepackage{hyperref}
\usepackage{psfrag}

\newcommand{\bq}{\begin{eqnarray}}
\newcommand{\eq}{\end{eqnarray}}

\newcommand{\ta}{\theta_1}
\newcommand{\tb}{\theta_2}
\newcommand{\ga}{\gamma_a}
\newcommand{\gb}{\gamma_b}

\newcommand{\cph}{c_\phi}
\begin{document}
\topmargin 0mm
\oddsidemargin 0mm

\begin{titlepage}
\begin{center}
\hfill USTC-ICTS-13-12\\
\vspace{2.5cm}
%\begin{center}

{\Large \bf CP mixed property of the Higgs-like particle in the decay channel $h\to Z Z^*\to 4l$}
\vspace{10mm}

{Yi Sun$^{*}$, Xian-Fu Wang$^{\dagger}$, Dao-Neng Gao$^{\ddagger}$}
\vspace{4mm}\\{\it\small
Interdisciplinary Center for Theoretical Study,
University of Science and Technology of China, Hefei, Anhui 230026 China}\\

\end{center}
\vspace{10mm}

\begin{abstract}
\noindent
Current experiments do not support, as ATLAS and CMS collaborations at the Large Hadron Collider reported, that the Higgs-like resonance discovered in July 2012 is a pure CP-odd state. We examine a general $hZZ$ vertex which contains CP-even and CP-odd couplings, by studying the process $h\to ZZ^*\to l_1^+l_1^-l_2^+l_2^-$ with $l_1,\, l_2= e$ or $\mu$, to explore the CP mixed property of the Higgs-like particle. One momentum asymmetry and two angular asymmetries have been analyzed in order to reveal the difference from different CP-couplings. Our study shows that these asymmetries could be interesting observables in the future precise experiments.
\end{abstract}

\vfill
\noindent
$^{*}$ E-mail:~sunyi@mail.ustc.edu.cn\\
\noindent
$^{\dagger}$ E-mail:~wangxf5@mail.ustc.edu.cn\\
$^{\ddagger}$ E-mail:~gaodn@ustc.edu.cn\\\noindent
\end{titlepage}
\newpage
\section{Introduction}
After more than forty years' efforts, a Higgs-like particle at around 125 GeV was finally found in July 2012 thanks to the hard work of ATLAS \cite{atlas} and CMS \cite{cms} Collaborations at the Large Hadron Collider (LHC).
In order to identify the new resonance with the standard model (SM) Higgs boson, one has to confirm its fundamental quantum numbers, e.g. spin and parity.
The authors of the literatures \cite{pseudoscalartheory, relevance1, relevance2, maltoni, anderson, choi03, boughezal12, godbole2007} have provided a useful method to discriminate between pure scalar and pure pseudoscalar state for the new particle.
The observations of the $h\to ZZ^*\to 4l$ and $h\to WW^*\to l\nu l\nu$ channels in the ATLAS \cite{Atlasreportzz, Atlasreport} and CMS \cite{CMSreportzz, CMSreport} show that the SM Higgs boson is consistent with the experiments and a pure CP-odd Higgs is disfavored.
However, it is also possible that the new particle, besides owning the similar coupling structures as the SM Higgs boson, is affected by the new physics (NP), for example, the CP violating NP which will introduce CP-odd component into the SM Higgs boson. In such scenario, the Higgs-like resonance should be a CP mixed state, rather than a CP eigenstate.
Thus it is of interest to explore the CP mixed property of the Higgs-like particle both to increase our understanding of the Higgs physics and to search for the NP beyond the SM.

Some authors \cite{relevance1, relevance2, maltoni, anderson, choi03, boughezal12, godbole2007} have studied the decay channel $h\to ZZ^*\to l_1^+l_1^- l_2^+l^-_2$
with $l_1$, $l_2=e$ or $\mu$
%and one of the intermediate $Z$ boson is on-shell while another one is off-shell
and investigated the effects of the CP-even and CP-odd $hZZ$ couplings.
For example, the authors of the literature \cite{boughezal12} defined a momentum asymmetry to study the $q^2$ distribution in the differential decay rate of $h\to ZZ^*\to 4l$ for various $hZZ$ coupling structures, where $q^2$ is the momentum square of the virtual intermediate gauge boson $Z$.
Godbole et al. \cite{godbole2007} constructed some angular asymmetries in this channel to directly probe the ratio of the CP-even and CP-odd components.

Since some events in this channel have been collected and well analyzed experimentally \cite{Atlasreportzz, CMSreportzz}, we further extend previous works \cite{boughezal12, godbole2007} and analyze a momentum asymmetry and two angular asymmetries to show the difference from different CP-couplings.
The newly reported mass value of the Higgs boson candidate from \cite{atlas, cms} will be used in our calculation.
These asymmetries can clearly show the ratios of different couplings in the general $hZZ$ vertex, in other words, the CP mixed property of the Higgs-like particle, and therefore might be interesting observables in the future precise experiments to explore the NP in the Higgs sector.

The paper is organized as follows.
In section 2, the amplitude of the process $h\to ZZ^*\to 4l$ is evaluated, and the expressions of the differential decay rate and some asymmetries are presented. Section 3 is our numerical analysis and some discussions on phenomenology. Finally, we give a summary in section 4.
\section{Amplitude of $h\to Z Z^*\to 4l$}

\begin{figure}
 \begin{center} \includegraphics[width=10cm]{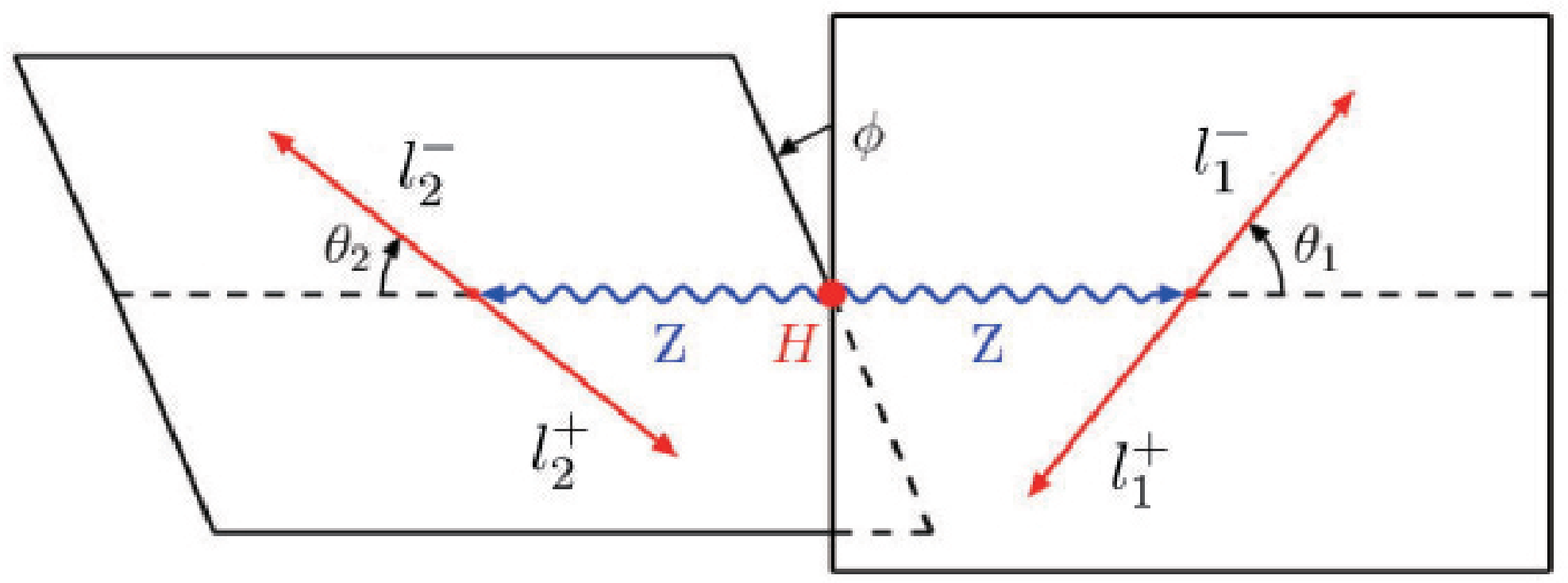}\end{center}
  \caption{The definitions of variables for $h\to ZZ^* \to 4l$}\label{tree}
\end{figure}

The process $h\to Z Z^*\to 4l$ is given in figure \ref{tree} where the Higgs boson decays to two gauge bosons $Z(p)$ and $Z^*(q)$, one of which is real with momentum square $p^2$ and the other is virtual with momentum square $q^2$, then each of them decays to a lepton pair.
The polar angles of lepton pairs $l_1^+$$l_1^-$ and $l_2^+$$l_2^-$ in the rest frame of their parent gauge boson $Z$ are denoted as $\theta_1$ and $\theta_2$, respectively, and the azimuthal angle of the
planes formed by each lepton pair and their parent gauge boson in the Higgs rest frame is denoted as $\phi$.
Some events in this process have been collected in the experiments.
ATLAS \cite{Atlasreportzz} selects the events satisfying $\sqrt{p^2}\in [50,106] \,\text{GeV}$ and $\sqrt{q^2}\in[12,115] \,\text{GeV}$ while for CMS \cite{CMSreportzz}, $\sqrt{p^2}\in [40,120] \,\text{GeV}$ and $\sqrt{q^2}\in[12,120] \,\text{GeV}$.
For this reason, to perform a complete theoretical analysis, one should regard $p^2$ as a variable. However, the phase space $[m_Z-\Gamma_Z, m_Z+\Gamma_Z]$ for $\sqrt{p^2}$ gives the dominant contribution, so we will take $p^2=m_Z^2$ throughout this paper for simplicity.
On the other hand, we choose $\sqrt{q^2}\in [12\,\text{GeV}, m_h-m_Z]$ in our calculation where the upper bound corresponds to the kinematically allowed maximum and the lower bound is set according to the experiments.
Since the $\tau^\pm$ pair in the final states is not easy to be observed, we discuss only, as the experiments did, four channels $h\to Z Z^*\to e^+e^-e^+e^-/e^+e^-\mu^+\mu^-/\mu^+\mu^-e^+e^-/\mu^+\mu^-\mu^+\mu^-$. Additionally, the masses of the light leptons  $m_e$ and $m_\mu$ in the final states only give the next leading order $O(\frac{m_l}{m_Z})$ contribution and hence will be neglected.

The general form of the effective $hZZ$ vertex, including both CP-even and CP-odd couplings, can be written as \cite{godbole2007}
\begin{eqnarray}
\mathcal{V}_{hZZ}=\frac{igm_Z}{\cos\theta_w}(a\,g_{\mu\nu}+b \frac{k_{1\mu} k_{2\nu}}{m_Z^2}+c~\epsilon_{\mu\nu\alpha\beta} \frac{k_1^\alpha k_2^\beta}{m_Z^2})\label{lag},
\end{eqnarray}
where $k_1=p+q$, $k_2=p-q$, $\theta_w$ is the Weinberg mixing angle, $g$ is the coupling constant of $SU(2)_L$ and $\epsilon_{\mu\nu\alpha\beta}$ is the totally antisymmetric tensor with $\epsilon_{0123}=1$.
The tree level SM coupling corresponds to $a=1$, $b=c=0$ and the SM Higgs boson is a pure CP-even state.
Although the parameters $a$ and $b$ will be corrected by the loop effect in the SM, we will identify $a=1$, $b=c=0$ as the SM case, since the contribution of the radiative corrections is small \cite{correction}.
The second term, another kind of CP-even coupling, is a high dimension operator which might be generated from the loop diagrams or heavy degrees of freedom.
The third term will introduce a CP-odd coupling and violate CP symmetry.
These three terms will be referred henceforth as $0_m^+$, $0_h^+$ and $0^-$ cases, respectively.
In general, when we go beyond the SM, parameters $a$, $b$ and $c$ can be complex.
However, since their imaginary parts might destroy the hermiticity of the effective theory \cite{godbole2007} and the process is dominated by the tree level in the SM, we will assume in this work, for simplicity, the parameters $a$, $b$ and $c$ in eq. (\ref{lag}) are all real.

The differential decay rate of the process $h\to ZZ^*\to 4l$ can be expressed as
\begin{eqnarray}
\frac{d^4\Gamma}{dc_{\ta}\, dc_{\tb} d\phi dq^2}
&=&\frac{p^2q^2}{2^{13}\pi^5m_h}\beta(a_1^2+v_1^2)(a_2^2+v_2^2)
BW(p^2)BW(q^2)\{\nonumber\\&&
a^2 (s_{\ta}^2 s_{\tb}^2
- \frac{1}{2\gamma_a} s_{2\ta} s_{2\tb} \cph
+ \frac{1}{2\gamma_a^2}
((1+c_{\ta}^2)(1+c_{\tb}^2)
+s_{\ta}^2 s_{\tb}^2 c_{2\phi}) \nonumber\\
&& -\frac{2\eta_1 \eta_2}{\gamma_a}
(s_{\ta} s_{\tb} \cph - \frac{1}{\gamma_a} c_{\ta} c_{\tb}) )
\nonumber \\
&+& b^2 \frac{\gamma_b^4}{\gamma_a^2} \, x^2
\, s_{\ta}^2 s_{\tb}^2 \nonumber \\
&+&
c^2 \frac{\gamma_b^2}{\gamma_a^2} \, 4 x^2 \,
( 1 + c_{\ta}^2 c_{\tb}^2 - \frac{1}{2} s_{\ta}^2 s_{\tb}^2(1 +
c_{2\phi}) +2\eta_1\eta_2 c_{\ta} c_{\tb} )\nonumber \\
&-& 2a b \frac{\gamma_b^2}{\gamma_a^2} \, x
\,(-\gamma_a s_{\ta}^2 s_{\tb}^2 + \frac{1}{4}s_{2\ta}s_{2\tb} \cph
+\eta_1 \eta_2 s_{\ta} s_{\tb}\cph ) \nonumber\\
&-& 2a c \frac{\gamma_b}{\gamma_a} \, 2x \,
s_{\ta} s_{\tb} s_{\phi}
( -c_{\ta} c_{\tb} + \frac{s_{\ta} s_{\tb} \cph}{\gamma_a} -\eta_1 \eta_2)
\nonumber\\
&+ &
2b c
\frac{\gamma_b^3}{\gamma_a^2}\, 2x^2 \, s_{\ta} s_{\tb} s_\phi
( c_{\ta} c_{\tb} + \eta_1 \eta_2)\},
\label{mmstar}
\end{eqnarray}
which can be obtained by setting the parameters $a$, $b$ and $c$ in the literature \cite{godbole2007} real.
The notations $s_{\theta_i}$, $c_{\theta_i}$ and $\cph$ denote $\sin{\theta_i}$, $\cos{\theta_i}$ and $\cos\phi$, respectively.
The $\eta_i$, in terms of the vector and axial vector couplings of $Zl^+l^-$, is
\begin{eqnarray}
\eta_i=\frac{2a_{l_i}v_{l_i}}{a_{l_i}^2+v_{l_i}^2} ~~~~\text{with}~~~~~~ a_{l_i}=T^3_{l_i} ~~~~~~\text{and}~~~~~~ v_{l_i}=T^3_{l_i}-2Q_{l_i} \sin^2{\theta_w},
\end{eqnarray}
where $T^3_{l_i}$ is the third component of the weak isospin and $Q_{l_i}$ is, in unit of e, the electric charge of the lepton $l_i^-$.
Some variables $\gb=\frac{m_h^2}{2\sqrt{p^2q^2}}\beta$, $\ga^2=\gb^2+1$ and $x=\frac{\sqrt{p^2q^2}}{m_Z^2}$ are introduced in eq. (\ref{mmstar}) while $\beta$ and $BW(s)$ are defined as
\begin{eqnarray}
\beta&=&\frac{\sqrt{(m_h^2-p^2-q^2)^2-4p^2q^2}}{m_h^2},\nonumber\\
BW(s)&=&\frac{1}{(s-m_Z^2)^2+m_Z^2\Gamma_Z^2}.
\end{eqnarray}
For the real gauge boson $Z$, $BW(s)=\frac{1}{m_Z^2\Gamma_Z^2}$.

Integrating the differential decay rate over all of the angular variables, one obtains
\begin{eqnarray}
\frac{d\Gamma}{dq^2}=\frac{p^2 q^2}{2304 \pi^5m_h}\beta (a_1^2+v_1^2) (a_2^2+v_2^2)BW(p^2)BW(q^2)(2 a^2+8(c x \gb)^2+(a \ga+b x \gb^2)^2)\label{differentialwidth},
\end{eqnarray}
and the decay rate of this process reads
\begin{eqnarray}
\Gamma=\int_{(12\,\text{GeV})^2}^{(m_h-m_Z)^2}\frac{d\Gamma}{dq^2}dq^2.
\end{eqnarray}
The lepton masses have been neglected in the above equations due to $m_e, m_\mu\ll m_Z$.
Actually, if we study the decay channel $h\to ZZ^*\to e^+e^-e^+e^-$ or $\mu^+\mu^-\mu^+\mu^-$, additional symmetry factor $\frac{1}{4}$ should be included. However, all of the asymmetries discussed in the present work are ratios and independent on the overall factor, one can forget the symmetry factor in the calculation.

The $q^2$ distributions in the differential decay rate of $h\to ZZ^*\to 4l$ for different $hZZ$ couplings could be different.
The authors of the literature \cite{boughezal12} constructed an interesting observable called high-low asymmetry, which is defined as
\begin{eqnarray}
\mathcal{A}_{HL}(q^2_{high}, q^2_{mid}, q^2_{low})=
\frac{\int^{q^2_{high}}_{q^2_{mid}}\frac{d\Gamma}{dq^2}dq^2
-\int^{q^2_{mid}}_{q^2_{low}}\frac{d\Gamma}{dq^2}dq^2}
{\int^{q^2_{high}}_{q^2_{mid}}\frac{d\Gamma}{dq^2}dq^2
+\int^{q^2_{mid}}_{q^2_{low}}\frac{d\Gamma}{dq^2}dq^2},\label{AHL}
\end{eqnarray}
to probe such difference.
Comparing with ATLAS \cite{Atlasreportzz2012} and CMS \cite{CMSreportzz2012} data, they found the spin-2 Higgs is disfavored by the experiments.
We extend their analysis to the $0^+_h$ case, which was not considered in the literature \cite{boughezal12}, and reveal some interesting phenomena. This will be discussed in the next section.

In order to directly probe various couplings in the general $hZZ$ vertex, we shall study, according to ref. \cite{godbole2007}, two angular asymmetries, which might be very interesting observables in the future experiments.
The first one is
\begin{eqnarray}
A_{1}(q^2)&=&
(\int^{\pi}_{\pi/2}\frac{d\Gamma}{dq^2d\phi}d\phi
+\int^{2\pi}_{3\pi/2}\frac{d\Gamma}{dq^2d\phi}d\phi
-\int^{\pi/2}_{0}\frac{d\Gamma}{dq^2d\phi}d\phi
-\int^{3\pi/2}_{\pi}\frac{d\Gamma}{dq^2d\phi}d\phi)/\frac{d\Gamma}{dq^2}\nonumber\\
&=&\frac{4acx\gamma_b}{\pi(2 a^2+8(c x \gb)^2+(a \ga+b x \ga\gb^2)^2)}.
\end{eqnarray}
It is easily found that such asymmetry selects terms containing $\sin2\phi$ and is proportional to $ac$.
The corresponding integrated asymmetry over $q^2$ is
\begin{eqnarray}
\mathcal{A}_{1}(q^2_{low},q^2_{high})=\frac{\int_{q^2_{low}}^{q^2_{high}}   4q^2\beta BW(q^2)a c x \gamma_b dq^2}{\int_{q^2_{low}}^{q^2_{high}} \pi q^2\beta BW(q^2)(2 a^2+8(c x \gb)^2+(a \ga+b x \gb^2)^2)dq^2}\label{A1}.
\end{eqnarray}
%Obviously, the non-vanishing $\mathcal{A}_{1}$ means that the Higgs-like particle is CP-violating.
By observing these asymmetries in the experiments, one can get information about the CP-odd term and have a more profound understanding about the CP mixed property of the Higgs-like particle.

The other angular asymmetry can be defined as
\begin{eqnarray}
A_{2}(q^2)&=&(\int^{\pi}_{0}\int^{\pi/2}_{0}\int^{\pi/2}_{0}
 +\int^{\pi}_{0}\int^{\pi}_{\pi/2}\int^{\pi}_{\pi/2}
 +\int^{2\pi}_{\pi}\int^{\pi/2}_{0}\int^{\pi}_{\pi/2}
 +\int^{2\pi}_{\pi}\int^{\pi}_{\pi/2}\int^{\pi/2}_{0}\nonumber\\
 &&-\int^{\pi}_{0}\int^{\pi/2}_{0}\int^{\pi}_{\pi/2}
 -\int^{\pi}_{0}\int^{\pi}_{\pi/2}\int^{\pi/2}_{0}
 -\int^{2\pi}_{\pi}\int^{\pi/2}_{0}\int^{\pi/2}_{0}
 -\int^{2\pi}_{\pi}\int^{\pi}_{\pi/2}\int^{\pi}_{\pi/2})\nonumber\\
 &&\frac{d\Gamma}{dq^2d\cos\theta_1d\cos\theta_2d\phi}d\cos\theta_1d\cos\theta_2d
 \phi/\frac{d\Gamma}{dq^2}\nonumber\\
&=&\frac{2cx\gamma_b(a \ga+b x \gb^2)}{\pi(2 a^2+8(c x \gb)^2+(a \ga+b x \gb^2)^2)}.
\label{A2}
\end{eqnarray}
It is seen that such asymmetry selects terms containing $\cos\theta_1\cos\theta_2\sin\phi$ and is contributed by terms containing $ac$ or $bc$.
The corresponding integrated asymmetry over $q^2$ is
\begin{eqnarray}
\mathcal{A}_{2}(q^2_{low},q^2_{high})=\frac{\int_{q^2_{low}}^{q^2_{high}} 2q^2\beta BW(q^2) c x \gamma_b(a \ga+b x \gb^2)dq^2}{\int_{q^2_{low}}^{q^2_{high}} \pi q^2\beta BW(q^2)(2 a^2+8(c x \gb)^2+(a \ga+b x \gb^2)^2)dq^2}\label{A2}.
\end{eqnarray}

The observables $\mathcal{A}_1$ and $\mathcal{A}_2$ are not totally independent in our formula. After performing a combined analysis of them, one can construct an interesting quantity as
\begin{eqnarray}
\mathcal{R}=\frac{\mathcal{A}_2(q^2_{low},q^2_{high}) g_1(q^2_{low},q^2_{high})}{\mathcal{A}_1(q^2_{low},q^2_{high})g_3(q^2_{low},q^2_{high})}
-\frac{g_2(q^2_{low},q^2_{high})}{g_3(q^2_{low},q^2_{high})},\label{deff}
\end{eqnarray}
with
\begin{eqnarray}
g_1(q^2_{low},q^2_{high})&=&\int_{q^2_{low}}^{q^2_{high}}2q^2\beta x \gamma_bBW(q^2)dq^2,~~
g_2(q^2_{low},q^2_{high})=\int_{q^2_{low}}^{q^2_{high}}q^2\beta x \gamma_a\gamma_bBW(q^2)dq^2,~~\nonumber\\
g_3(q^2_{low},q^2_{high})&=&\int_{q^2_{low}}^{q^2_{high}}q^2\beta x^2 \gamma_b^3BW(q^2)dq^2.
\end{eqnarray}
$\mathcal{R}\equiv\frac{b}{a}$, by definition, is a constant and proportional to $b$ in our formula and has no any relationship with the momentum region\footnote{Generally, $a,\,b$ and $c$ should be momentum dependent. Since such dependence is origin from integrating over the degree of freedom in the high energy scale, it is reasonable to suppose these parameters are constant in the electroweak scale.} appearing on the right-hand side of eq. (\ref{deff}).
For this reason, the studies of this quality in the experiments are very important.
First, the parameter $\mathcal{R}$ is proportional to $b$ and hence reflects the magnitude of the $0_h^+$ component of the Higgs-like particle.
If $\mathcal{R}$ is close to zero, which means $b \,\sim\, 0$, the $0_h^+$ component of the Higgs-like particle is negligible while the large $\mathcal{R}$ means that the $0_h^+$ component is important.
Second, it is also interesting to study such quantity in different momentum regions in the experiments and examine whether it is a constant or not.
If the answer is negative, the most possible reason is that the momentum-dependence of the parameters $a$, $b$ and $c$ cannot be neglected.
%, which hints that the energy scale of the new physics is not so high as expected.
Of course, it is also not an easy work to fix that $\mathcal{R}$ is not a constant experimentally.

\section{Analysis}
\subsection{High-low asymmetry}
\begin{figure}
  \begin{center} \includegraphics[width=10cm]{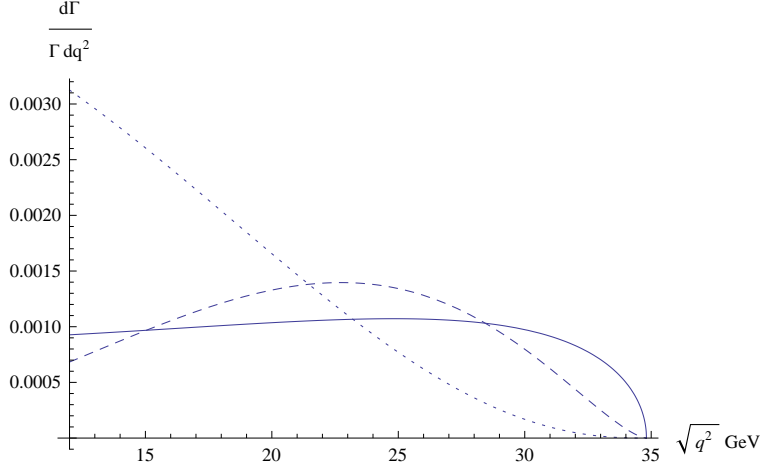}\end{center}
  \caption{The normalized differential decay rate of $h\to ZZ^*\to 4l$ for various couplings. The real line shows the behavior of the SM Higgs, the dotted line corresponds to the $0_h^+$ case and the dashed line the $0^-$ case.}\label{ahldiff}
\end{figure}

The $q^2$ distributions in the normalized differential decay rate of $h\to ZZ^*\to 4l$ for various couplings have been plotted in figure \ref{ahldiff}.
We found that the behavior of the $0_h^+$ Higgs boson is different from $0_m^+$ and $0^-$ cases.
Such difference can also be seen by studying the corresponding high-low asymmetries. Here we choose the high energy region as $[22\,\text{GeV}, 32\,\text{GeV}]$ and the low energy region $[12\,\text{GeV}, 22\,\text{GeV}]$, which means the parameters in eq. (\ref{AHL}) are set to be
$\sqrt{q_{high}^2}=32\,\text{GeV}$, $\sqrt{q^2_{mid}}=22\,\text{GeV}$ and $\sqrt{q^2_{low}}=12\,\text{GeV}$.

After the successful running of the LHC in 2011 and 2012, some events in the decay channel $h\to ZZ^*\to 4l$ have been collected, therein ATLAS \cite{Atlasreportzz} observed 7.8 events at the high energy region and 8.1 events at the low energy region while CMS \cite{CMSreportzz} reported 7.6 and 5.1.
According to the eq. (\ref{AHL}), one can obtain the high-low asymmetries as
\begin{eqnarray}
\mathcal{A}_{HL}^{\text{ATLAS}}=-0.02\pm0.094,~~~~~~
\mathcal{A}_{HL}^{\text{CMS}}=0.19\label{data}.\end{eqnarray}
The uncertainty on the $\mathcal{A}_{HL}^{\text{ATLAS}}$ is evaluated according to the uncertainty on the experiment data as given in the literature \cite{Atlasreportzz} whereas for the CMS \cite{CMSreportzz}, the uncertainty is not available.
The high-low asymmetries for various couplings in our formula can be calculated as
\begin{eqnarray}
\mathcal{A}_{HL}^{0^+_m}=0.23,~~\mathcal{A}_{HL}^{0^+_h}=-0.45,
~~\mathcal{A}_{HL}^{0^-}=0.20,
\end{eqnarray}
which correspond to the SM, $0^+_h$, $0^-$ cases, respectively.
One can find that $\mathcal{A}_{HL}^{0^+_h}$ is notably different from $\mathcal{A}_{HL}^{0^+_m}$ and $\mathcal{A}_{HL}^{0^-}$ and also deviates from the experiment values $\mathcal{A}_{HL}^{\text{ATLAS}}$ and $\mathcal{A}_{HL}^{\text{CMS}}$ as given in eq. (\ref{data}).
On the other hand, the high-low asymmetry value of the SM Higgs boson $\mathcal{A}_{HL}^{0^+_m}$ is consistent with that of the ATLAS $\mathcal{A}_{HL}^{\text{ATLAS}}$ and CMS $\mathcal{A}_{HL}^{\text{CMS}}$.
%\textbf{However, it seems that, according to the central value of $\mathcal{A}_{HL}^{\text{ATLAS}}$ in eq. (\ref{data}), some new components might be needed.}
Since the behaviors of the $0^+_m$ and $0^-$ Higgs are similar in figure \ref{ahldiff}, such asymmetry cannot give any useful constraint on the CP-odd component.

\subsection{Angular asymmetries}
\begin{figure}
  \begin{center}
  \includegraphics[width=10cm]{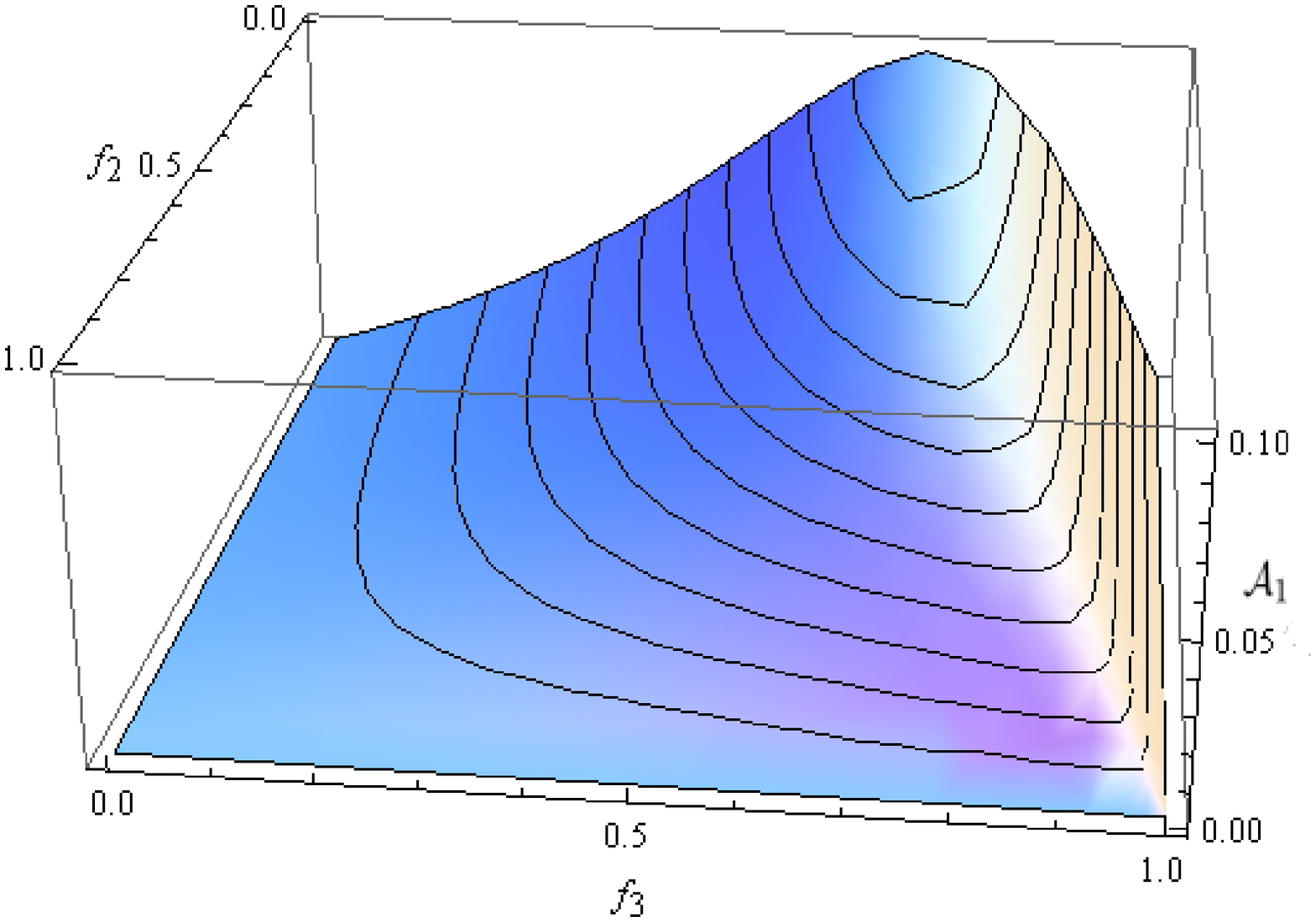}\end{center}
  \caption{The integrated asymmetry $\mathcal{A}_1$ with $\sqrt{q^2_{low}}=12\,\text{GeV}$ and $\sqrt{q^2_{high}}=(m_h-m_Z)$ in the full parameter space of $f_2$ and $f_3$.}\label{a1int}
\end{figure}
\begin{figure}
  \begin{center}
  \includegraphics[width=10cm]{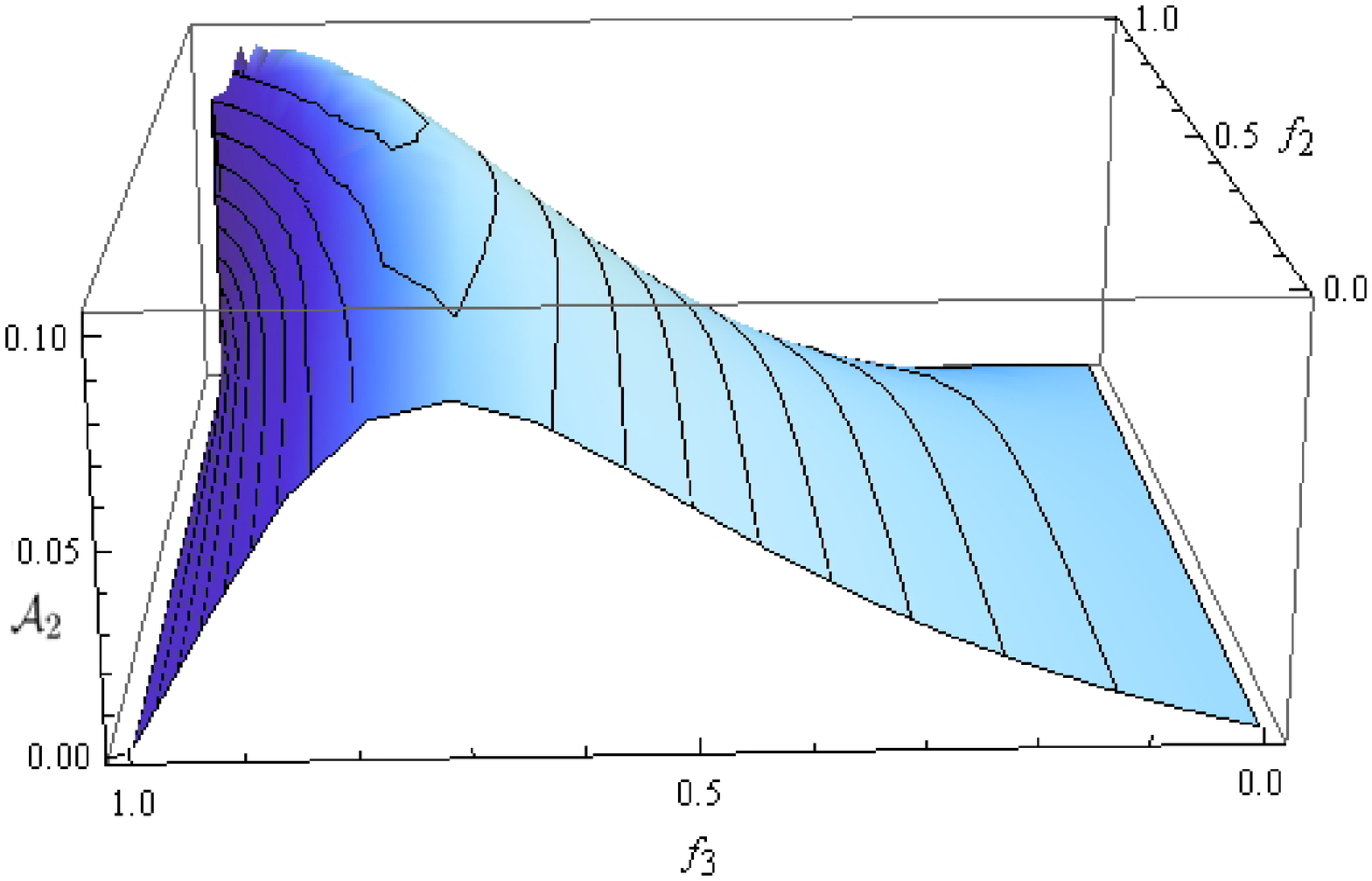}\end{center}
  \caption{The integrated asymmetry $\mathcal{A}_2$ with $\sqrt{q^2_{low}}=12\,\text{GeV}$ and $\sqrt{q^2_{high}}=(m_h-m_Z)$ in the full parameter space of $f_2$ and $f_3$.}\label{a2int}
\end{figure}
\begin{figure}
  \begin{center} \includegraphics[width=10cm]{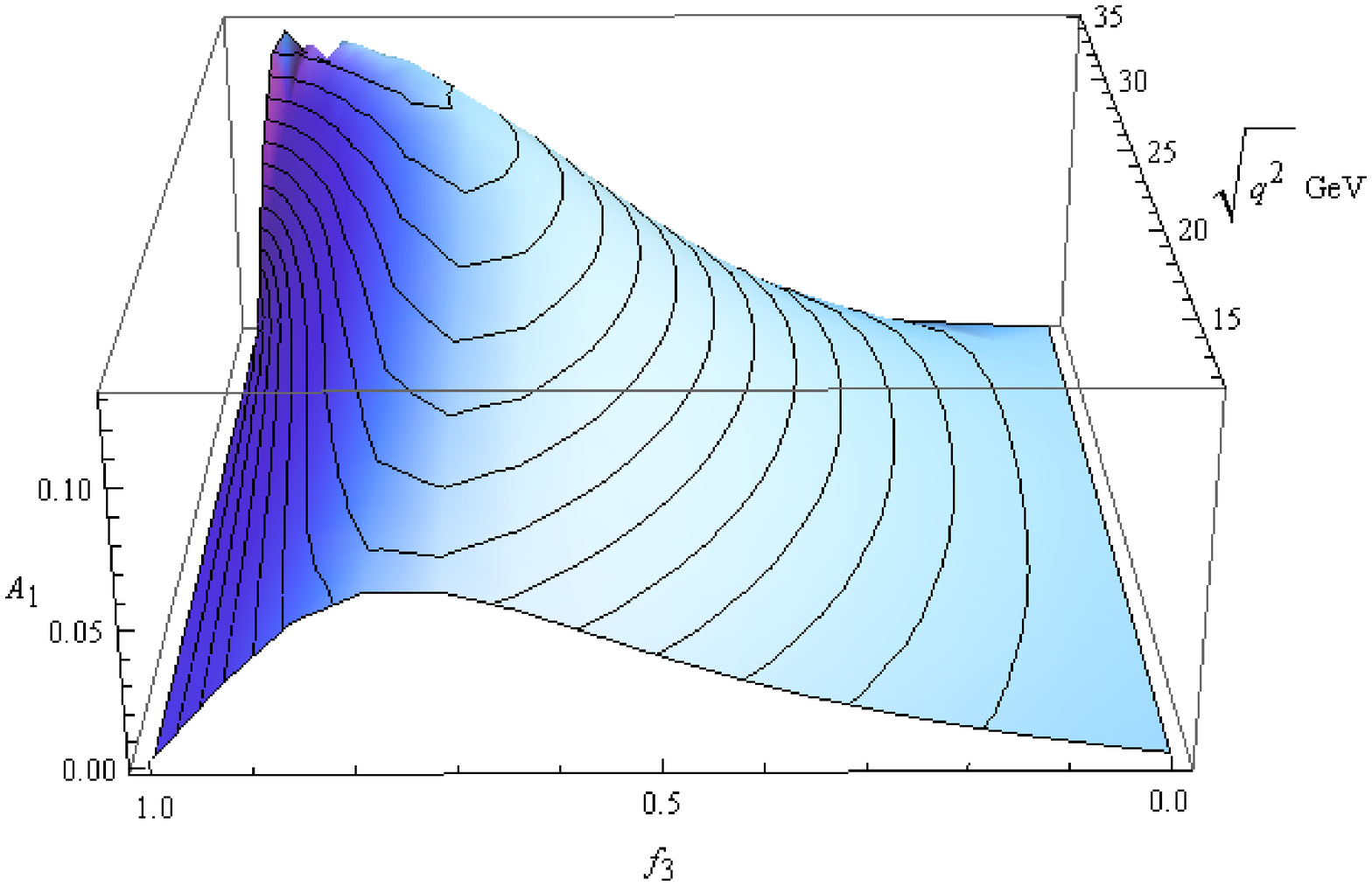}\end{center}
  \caption{The angular asymmetry $A_1$ with the change of $f_3$ and the invariant mass of the lepton pair $\sqrt{q^2}$ where $f_2$ is set to be zero.}\label{a1f20}
\end{figure}
\begin{figure}
  \begin{center} \includegraphics[width=10cm]{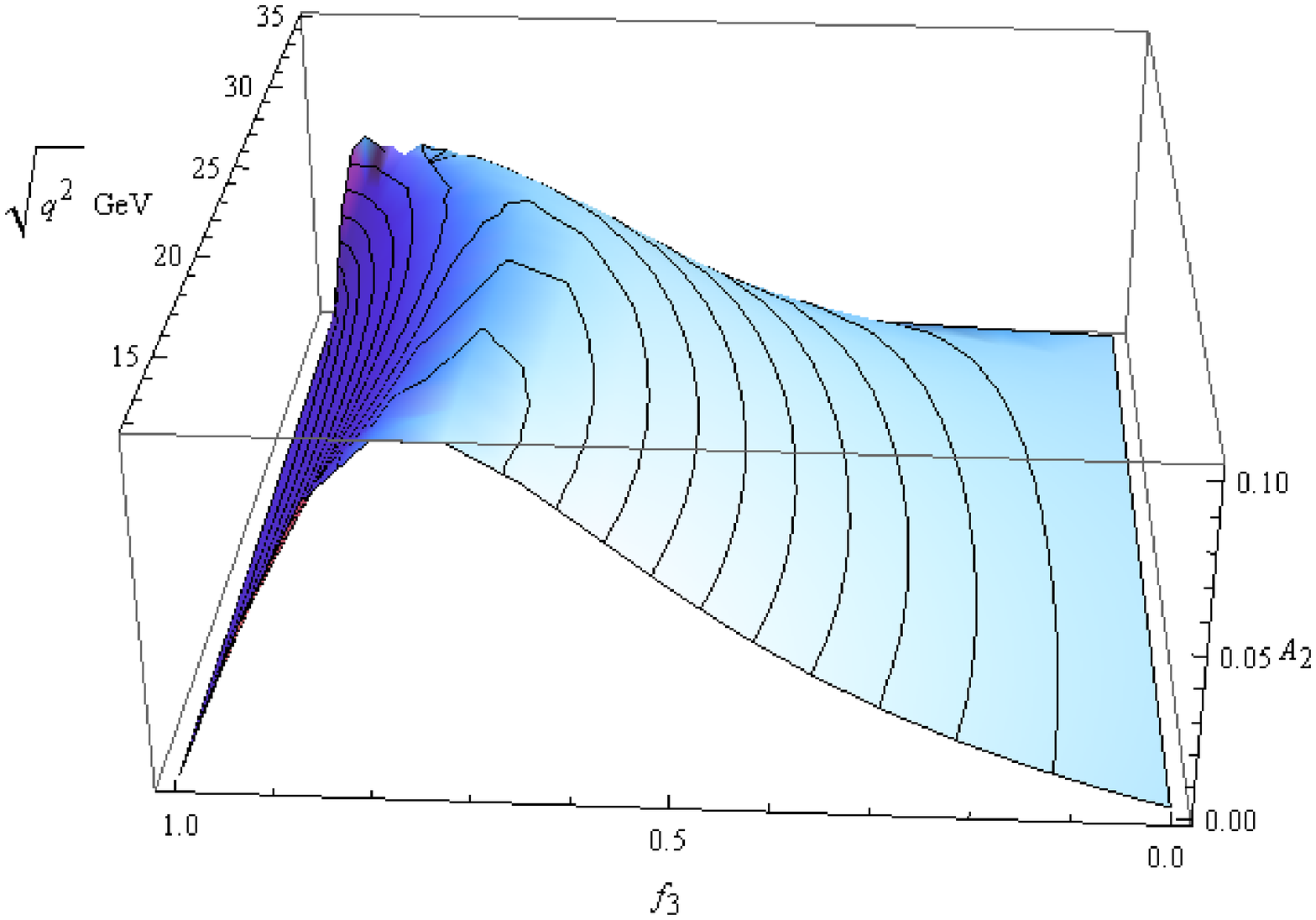}\end{center}
  \caption{The angular asymmetry $A_2$ with the change of $f_3$ and the invariant mass of the lepton pair $\sqrt{q^2}$ where $f_2$ is set to be zero.}\label{a2f20}
\end{figure}

In order to analyze the CP mixed property of the Higgs-like particle in the full parameter space of $a$, $b$ and $c$, we define two useful variables as
\begin{eqnarray}
f_2=\frac{b}{a+b},~~~~~~~~~f_3=\frac{c}{a+c},
\end{eqnarray}
where $0\leq (f_2, f_3)\leq 1$. $f_2=f_3=0$ is associate with the $0^+_m$ case, $f_2=1$ and $f_3=0$ corresponds to the $0^+_h$ case while $f_3=1$ and $f_2=0$ the $0^-$ case.
The integrated asymmetries $\mathcal{A}_1$ and $\mathcal{A}_2$ in the full parameter space of $f_2$ and $f_3$ have been plotted in figures \ref{a1int} and \ref{a2int}, respectively, where the variables of integration in eqs. (\ref{A1}) and (\ref{A2}) are set to be $\sqrt{q^2_{low}}=12\,\text{GeV}$ and $\sqrt{q^2_{high}}=(m_h-m_Z)$ with $m_h = 125 \text{GeV}$.
It is found that the behaviors of $\mathcal{A}_1$ and $\mathcal{A}_2$ are similar.
They both reach the maximal value, which could be up to $10^{-1}$, when $f_3\sim 0.7$, and can be higher than $3\%$ in most of the parameter space.
Actually, these integrated asymmetries $\mathcal{A}_1$ and $\mathcal{A}_2$, as a function of $c/a$, with $b=0$ and $m_h=150~\text{GeV}$, have been investigated in the literature \cite{godbole2007}.

Taking a close look at figures \ref{a1int} and \ref{a2int}, one can find that the integrated asymmetries, when $f_2<0.7$, are not sensitive to the change of $f_2$.
We therefore choose $f_2=0$ and plot the behaviors of the differential asymmetries $A_1$ and $A_2$ with the change of $f_3$ and $\sqrt{q^2}$, as shown in figures \ref{a1f20} and \ref{a2f20}.
The maximal values of $A_1$ and $A_2$ are $11\%$ and $13\%$, respectively, and these asymmetries can be larger than $3\%$ in most of the parameter space.
We also notice that, if the NP effect is weak, e.g. $b, c \sim 0.1 a$, the angular asymmetries, only about $1\%$, are also hard to be observed and there will be a long way for us to understand the NP.
\section{Summary}
We have examined the general $hZZ$ vertex which contains the SM, high dimension CP-even and CP-odd coupling structures.
By studying the process $h\to ZZ^*\to l_1^+l_1^-l_2^+l_2^-$ with $l_1,\, l_2= e$ or $\mu$, we have analyzed the high-low momentum asymmetry and two angular asymmetries to explore the CP mixed property of the Higgs-like particle.
The newly reported mass value of the Higgs boson candidate from the LHC \cite{atlas, cms}, $m_h$=125 GeV, has been used in our calculation.

The studies of the high-low momentum asymmetry has shown that the $0^+_h$ Higgs is disfavored by the present data.
In order to further understand the nature of the CP-odd component, we have analyzed two angular asymmetries in the process $h\to ZZ^*\to 4l$.
The integrated angular asymmetries $\mathcal{A}_1$ and $\mathcal{A}_2$ could be over $3\%$ in most of the parameter space, and their maximal values can be up to $10^{-1}$.
Therefore these asymmetries are worth to be seriously considered in the future precise experiments and may provide us some important information on the CP mixed property of the new resonance.

\section*{Acknowledgements}
This work was supported in part by the NSF of China under Grant Nos. 11075149 and 11235010.

%%%% 参考文献排版格式：


\begin{thebibliography}{99}
\itemsep=-4pt plus.2pt minus.2pt  %% 调整参考文献条与条之间的间距
\small
\bibitem{atlas}
  ATLAS Collaboration,
  \emph{Observation of a new particle in the search for the Standard Model
  Higgs boson with the ATLAS detector at the LHC},
  \href{http://www.sciencedirect.com/science/article/pii/S037026931200857X}
  { \emph{Phys. Lett.} \textbf{B 716} (2012) 1}
  [\href{http://arxiv.org/abs/1207.7214}{arXiv:1207.7214}].
\bibitem{cms}
     CMS Collaboration, \emph{Observation of a new boson at a mass of 125 GeV with the CMS experiment at the LHC}, \href{http://www.sciencedirect.com/science/article/pii/S0370269312008581}
     {\emph{Phys. Lett.} \textbf{B 716} (2012) 30} [\href{http://arxiv.org/abs/1207.7235}{arXiv:1207.7235}]; \emph{Observation of a new boson with mass near 125 GeV in PP collisions at $\sqrt{s}$=7 TeV and 8 TeV}, \href{http://link.springer.com/article/10.1007\%2FJHEP06\%282013\%29081}
     {\emph{JHEP} \textbf{06} (2013) 081} [\href{http://arxiv.org/abs/1303.4571}{arXiv:1303.4571}].
\bibitem{pseudoscalartheory}
     Y. Gao, A. V. Gritsan, Z. Guo, K. Melnikov, M. Schulze, and N. V. Tran, \emph{Spin determination of single-produced resonances at hadron colliders},
     \href{http://prd.aps.org/abstract/PRD/v81/i7/e075022}
     {\emph{Phys. Rev.} \textbf{D 81} (2010) 075022}
     [\href{http://arxiv.org/abs/1001.3396}{arXiv:1001.3396}];
     S. Bolognesi, Y. Gao, A. V. Gritsan, K. Melnikov, M. Schulze, N. V. Tran and A. Whitbeck,
 \emph{Spin and parity of a single-produced resonance at the LHC}, \href{http://prd.aps.org/abstract/PRD/v86/i9/e095031}{\emph{Phys. Rev.} \textbf{D 86} (2012) 095031}
     [\href{http://arxiv.org/abs/1208.4018}{arXiv:1208.4018}].
\bibitem{relevance1}
T.~Plehn, D.~Rainwater and D.~Zeppenfeld,
   \emph{Determining the Structure of Higgs Couplings at the CERN Large Hadron Collider},
    \href{http://prl.aps.org/abstract/PRL/v88/i5/e051801}{\emph{Phys. Rev. Lett.}  {\bf 88} (2002) 051801}
    [\href{http://arxiv.org/abs/hep-ph/0105325}{hep-ph/0105325}];
C.~P.~Buszello, I.~Fleck, P.~Marquard and J.~J.~van der Bij,
    \emph{Prospective analysis of spin- and CP-sensitive variables in H $\to$ Z Z $\to l_1^+ l_1^- l_2^+ l_2-$ at the LHC},
    \href{http://link.springer.com/article/10.1140/epjc/s2003-01392-0#}
    {\emph{Eur. Phys. J.} {\bf C 32} (2004) 209}
    [\href{http://arxiv.org/abs/hep-ph/0212396}{hep-ph/0212396}];
  C.~P.~Buszello, P.~Marquard and J.~J.~van der Bij,
  \emph{On the determination of the structure of the scalar Higgs boson's
  couplings to vectorbosons},
  \href{http://arxiv.org/abs/hep-ph/0406181}{hep-ph/0406181};
  P. Niezurawski, A. F. Zarnecki and M. Krawczyk, \emph{Model-Independent Determination of CP Violation from Angular Distributions in Higgs Boson Decays to WW and ZZ at the Photon Collider}, \href{http://www.actaphys.uj.edu.pl/_old/vol36/abs/v36p0833.htm}
  {\emph{Acta Phys. Polon.} \textbf{B 36} (2005) 833}
  [\href{http://arxiv.org/abs/hep-ph/0410291}{hep-ph/0410291}].
\bibitem{relevance2}
D. Stolarski and R. Vega-Morales, \emph{Directly Measuring the Tensor Structure of the Scalar Coupling to Gauge Bosons}, \href{http://prd.aps.org/abstract/PRD/v86/i11/e117504}{\emph{Phys. Rev.} \textbf{D 86 } (2012) 117504}
[\href{http://arxiv.org/abs/1208.4840}{arXiv:1208.4840}];
A. Alves, \emph{Is the New Resonance Spin 0 or 2? Taking a Step Forward in the Higgs Boson Discovery}, \href{http://prd.aps.org/abstract/PRD/v86/i11/e113010}{\emph{Phys. Rev.} \textbf{D 86} (2012) 113010}
[\href{http://arxiv.org/abs/1209.1037}{arXiv:1209.1037}];
  A. Djouadi, R. M.Godbole, B. Mellado and K. Mohan, \emph{Probing the spin-parity of the Higgs boson via jet kinematics
in vector boson fusion}, \href{http://www.sciencedirect.com/science/article/pii/S0370269313003493}
{\emph{Phys. Lett.} \textbf{B 723} (2013) 307}
[\href{http://arxiv.org/abs/1301.4965}{arXiv:1301.4965}];
A. Djouadi and G. Moreau, \emph{The couplings of the Higgs boson and its CP properties from fits of the signal strengths and their ratios at the 7+8 TeV LHC}, \href{http://link.springer.com/article/10.1140/epjc/s10052-013-2512-9#}{ \emph{Eur. Phys. J.} \textbf{C 73} (2013) 2512}[\href{http://arxiv.org/abs/1303.6591}{arXiv:1303.6591}].
\bibitem{maltoni}
    P. Artoisenet et al., \emph{A framework for Higgs characterisation}, \href{http://link.springer.com/article/10.1007/JHEP11(2013)043#}{\emph{JHEP} \textbf{11} (2013) 043}
    [\href{http://arxiv.org/abs/1306.6464}{arXiv:1306.6464}].
\bibitem{anderson}
    Ian Anderson et al., \emph{Constraining anomalous HVV interactions at proton and lepton colliders}, \href{http://arxiv.org/abs/1309.4819}{arXiv:1309.4819}.
\bibitem{choi03}
    S. Y. Choi, D. J. Miller, M. M. Muhlleitner and P. M. Zerwas,
    \emph{Identifying the Higgs Spin and Parity in Decays to Z Pairs},
     \href{http://www.sciencedirect.com/science/article/pii/S037026930203191X}
    {\emph{Phys. Lett.} \textbf{B 553} (2003) 61} [\href{http://arxiv.org/abs/hep-ph/0210077v1}{hep-ph/0210077}].
\bibitem{boughezal12}
    R. Boughezal, T. J. LeCompte and F. Petriello, \emph{Single-variable asymmetries for measuring the Higgs' boson spin and CP properties}, \href{http://arxiv.org/abs/1208.4311}{arXiv:1208.4311}.
\bibitem{godbole2007}
    R. M. Godbole, D. J. Miller and M. M. M\"{u}hlleitner,
  \emph{Aspects of CP violation in the HZZ coupling at the LHC}, \href{http://iopscience.iop.org/1126-6708/2007/12/031}{\emph{JHEP} \textbf{12} (2007) 031} [\href{http://arxiv.org/abs/0708.0458}{arXiv:0708.0458}].
\bibitem{Atlasreportzz}
    ATLAS collaboration,
\emph{Measurements of the properties of the Higgs-like boson in the four lepton decay channel with the ATLAS detector using 25 fb-1 of proton-proton collision data}, \href{http://cds.cern.ch/record/1523699}{ATLAS-CONF-2013-013}.
\bibitem{Atlasreport}
     ATLAS collaboration,
    \emph{Evidence for the spin-0 nature of the Higgs boson using ATLAS data},
    \href{http://www.sciencedirect.com/science/article/pii/S0370269313006527}{\emph{Phys. Lett.} \textbf{B 726} (2013) 120}
    [\href{http://arxiv.org/abs/1307.1432}{arXiv:1307.1432}]; \emph{Study of the spin of the Higgs-like boson in the two photon decay channel using 20.7 fb-1 of pp collisions collected at $\sqrt{s} = 8$ TeV with the ATLAS detector}, \href{http://cds.cern.ch/record/1527124}{ATLAS-CONF-2013-029}; \emph{Measurements of the properties of the Higgs-like boson in the $WW^*\to l\nu l\nu$ decay channel with the ATLAS detector using 25 fb-1 of proton-proton collision data},    \href{http://cds.cern.ch/record/1527126}{ATLAS-CONF-2013-030}; \emph{Study of the spin properties of the Higgs-like particle in the $h\to WW^*\to e\nu\mu\nu$ channel with 21 fb-1 of $\sqrt{s} = 8$ TeV data collected with the ATLAS detector},
    \href{http://cds.cern.ch/record/1527127}{ATLAS-CONF-2013-031}.
\bibitem{CMSreportzz}
     CMS Collaboration, \emph{Properties of the Higgs-like boson in the decay $H$ to $ZZ$ to $4l$ in pp collisions at $\sqrt{s} =7$ and $8$ TeV}, \href{http://cds.cern.ch/record/1523767?ln=en}{CMS-PAS-HIG-13-002}.
\bibitem{CMSreport}
      CMS collaboration, \emph{Study of the Mass and Spin-Parity of the Higgs Boson Candidate via Its Decays to Z Boson Pairs}, \href{http://prl.aps.org/abstract/PRL/v110/i8/e081803}{\emph{Phys. Rev. Lett.} \textbf{110} (2013) 081803} [\href{http://arxiv.org/abs/1212.6639}{arXiv:1212.6639}]; \emph{Evidence for a particle decaying to W+W- in the fully leptonic final state in a standard model Higgs boson search in pp collisions at the LHC}, \href{http://cds.cern.ch/record/1523673?ln=en}{CMS-PAS-HIG-13-003}.
\bibitem{correction}
    B.A.~Kniehl, \emph{Radiative corrections for $H\to ZZ$ in the standard model},
    \href{http://www.sciencedirect.com/science/article/pii/055032139190126I}{
\emph{Nucl.~Phys.}~\textbf{B~352} (1991) 1};
    A.~Bredenstein, A.~Denner, S.~Dittmaier and M.~M.~Weber,
\emph{Precise predictions for the Higgs-boson decay $H \to W W / Z Z \to $ 4 leptons}, \href{http://prd.aps.org/abstract/PRD/v74/i1/e013004}{\emph{Phys. Rev.} \textbf{D 74} (2006) 013004}
 [\href{http://arxiv.org/abs/hep-ph/0604011}{hep-ph/0604011}].
\bibitem{Atlasreportzz2012}
    ATLAS collaboration, \emph{Observation of an excess of events in the search for the Standard Model Higgs boson in the $H\to ZZ^*\to 4l$ channel with the ATLAS detector}, \href{http://cds.cern.ch/record/1460411}{ATLAS-CONF-2012-092}.
\bibitem{CMSreportzz2012}
    CMS collaboration, \emph{Updated results on the new boson discovered in the search for the standard model Higgs boson in the ZZ to 4 leptons channel in pp collisions at $\sqrt {s} = $7 and 8 TeV}, \href{http://cds.cern.ch/record/1494488?ln=en}{CMS-PAS-HIG-12-041}.
\end{thebibliography}
\end{document}